# Heuristic approach to optimize the number of test cases for simple circuits

SM. Thamarai[1], K.Kuppusamy[2] and T. Meyyappan[3]

[1]Research scholar
`lotusmeys@yahoo.com`
[2] Associate Professor&Research Supervisor
`kkdiksamy@yahoo.com`
[3] Associate Professor
`meyslotus@yahoo.com`

Department of Computer Science and Engineering, Alagappa University,
Karaikudi-630003, Tamilnadu, South India.

## ABSTRACT

*In this paper a new solution is proposed for testing simple stwo stage electronic circuits. It minimizes the number of tests to be performed to determine the genuinity of the circuit. The main idea behind the present research work is to identify the maximum number of indistinguishable faults present in the given circuit and minimize the number of test cases based on the number of faults that has been detected. Heuristic approach is used for test minimization part, which identifies the essential tests from overall test cases. From the results it is observed that, test minimization varies from 50% to 99% with the lowest one corresponding to a circuit with four gates .Test minimization is low in case of circuits with lesser input leads in gates compared to greater input leads in gates for the boolean expression with same number of symbols. Achievement of 99% reduction is due to the fact that the large number of tests find the same faults. The new approach is implemented for simple circuits. The results show potential for both smaller test sets and lower cpu times.*

## 1. INRODUCTION

In recent years, the development of integrated circuit technology has accelerated rapidly; MSI and LSI techniques promise to make today's functional level devices tomorrow's(even today's) basic components. Accordingly, digital systems are built with more and more complexity; the fault testing and diagnosis of digital circuits becomes an important and indispensable part of the manufacturing process.

Performance, area, power and testing are some of the most important attributes of complex VLSI

systems[7]. With the current reduction in devise sizes it is possible to fit increasingly larger devices on to a single chip. As chip density increases the probability of defects occuring in a chip increases as well. The quality, reliability and cost of the product are directly proportional to the degree of testing the product[1]. Deterministic test generation algorithms for combinational circuits[10] and sequential circuits [8],[9],[11],[12],[13],[14],[15] have been used in the past, but the exection time are often long, due to the large number of backtracks that often occur.

The objective of this paper is to develop new test generation algorithm using heuristic optimization techniques. This algorithm reduces the number of test vector sets of a combinational logic circuit for fault detection and diagnosis respectively. With increasing integrated levels in today's VLSI chips, the complexity of testing them is also increasing. This is because the internal chip modules have become increasingly difficult to access. Testing costs have become a significant fraction of the total manufacturing cost. Hence there is a necessity to reduce the testing cost. The facter that has the biggest impact on testing cost of a chip is the time required to test it. This time can be decreased if the number of tests required to test the chip is reduced. So we simply need to advise a test set that is small in size[1].

## 1.1 Problem Statement

The problems solved in this paper are:
1. Finding a fault dictionary for all stuck at faults in combinational circuit.
2. Formulating an exact method for minimizing the given diagnostic test set
3. Calculate the execution time to overcome the computational limitations of the exact method of item 2.

## 1.2 Original contributions:

We have developed the test minimization algorithm to produce minimal test set for combinational circuits. This method has its foundations on these steps: 1) Identifying the distinguishable faults 2) Generating test cases for them 3) Minimizing the number of test cases. The steps 1 and 2, give us a non-exhaustive vector set, which on compaction will give a minimal test set. Using fault detection and location, we have modeled fault dictionary. Fault dictionary is a table. In this table rows identify the test number and column identifies the distinguishable faults. Heuristic approach is adopted to optimize the number of test cases. Simple two stage circuits and its Boolean expressions (as sum of product form) are experimented with the proposed method and test minimizing is found to be satisfactory.

The earlier research work on test set compaction is discussed in section 2. In section 3, the steps involved in heuristic approach for test minimization is discussed. In section 4, complete algorithm for test set compaction is presented. In section 5, outcomes of various simple circuits are tabulated and results are interpreted. In the final section, conclusions and future research directions are outlined.

## 2. BACKGROUND

This paper addresses the problem of minimal test pattern generation for simple combinational logic circuits only. However it should be noted that nearly all sequential logic, ie the circuit containing state holding elements (flip flops) are tested in a way that transforms their operation under test from sequential to combinational [2]. A number of basic analytic and heuristic methods are found in the literature namely, Fault Table method, Path Sensitizing method and Equivalent-Normal-Form(ENF) method, Karnaugh Map and tabular method, the ENF Karnaugh map method, the Boolean Difference method, and the SPOOF method.[3]

The *fault table method* [3] [4] is the most classic approach to the problem. It is completely general and always yields the minimum set of diagnostic tests. However it suffers from the fact that it requires the very large fault table to be constructed. To overcome the problem of not requiring the construction of very large fault table, the concept of path sensitizing is introduced.

A heuristic, systematic procedure derived from the concept of path sensitizing, known as *equivalent normal form(ENF) method[4]* is then introduced. Although this method has eliminated the two imperfections that the fault table and path sensitizing methods have, it introduces an unattractive new feature, the requirement of the cumbersome computation of a 'score function' for every literal in the ENF and the complemented ENF of the circuit.

*ENF Karnaugh map method* [6]is the combination of ENF method and the Karnaugh map and tabular method. Both the ENF method and ENF-Karnaugh map method do not guarantee minimal experiments, nor is there a guarantee that a set of sensitized paths can be found for every circuit.

The Boolean Difference method and the SPOOF methods [4] are two convenient general methods for deriving tests for detecting any single and/or multiple faults in any part of the circuit without using any fault tables or maps.

In Fixed Scheduled Fault Detection method, three tables are generated for the given combinational circuit (boolean expression in sum of products form). They are fault table, fault detection table and fault location table. Essential test set is found using a new algorithm[16] . It removes redundancy in test set by grouping test numbers detecting the same faults. It also removes redundant test numbers that are detecting the same fault. Test numbers detecting single faults alone are also collected as essential test numbers

## 3. HEURISTIC METHOD FOR TEST MINIMIZATION

In heuristic method, fault detection and test minimization consists of two phases. In the first phase fault dictionary alone is created. In the second phase number of tests is going to be minimized. In this paper diagnosing tree is created by dissecting the fault table matrix into two sub matrices based on essential test number. The test number is added to essential test set. Column numbers in these two matrices are added to the root node of the tree as right and left siblings. Left children contains fault-free output column numbers from the matrix (0s) and the right one contains faulty output column numbers from the matrix(1s).The process is repeated until both left and right children results in a single column number in them. Essential test set is found after removing redundant test numbers in it[4].

In this method, the economies that can be achieved by choosing each test input to be applied to the circuit on the basis of the outcomes of all previous tests in the schedule. The choice of test schedule depends upon the outcome of the individual tests in the sequence.  A convenient way to present such a sequence of tests and their outcomes is to use a diagnosing tree.

### 3.1 Diagnosing Tree

A diagnosing tree is a directed graph whose nodes are tests. The outgoing branches from a node represent the different outcomes of the particular test.  The diagnosing tree[3] is shown in figure 1,2 and 3.From this diagnosing tree, we see that the circuit fault-free if and only if the output sequence to the sequence (2,3,6,5) is (0,1,1,0), as indicated by the path of dark lines. Thus this tree can be simplified to the one shown in figure 2. Applying test 2,3,5,6 in any order will not shorten the length of the experiment. Another diagnosing tree for test set {2,3,5,6) is shown in figure 3, which still requires all four tests. Unlike fault detection, adaptive-scheduled fault location is general yields a shorter experiment. Using heuristic approach, the fault location of the circuit of figure 1, requires a minimal length of 4 is required for test minimization.

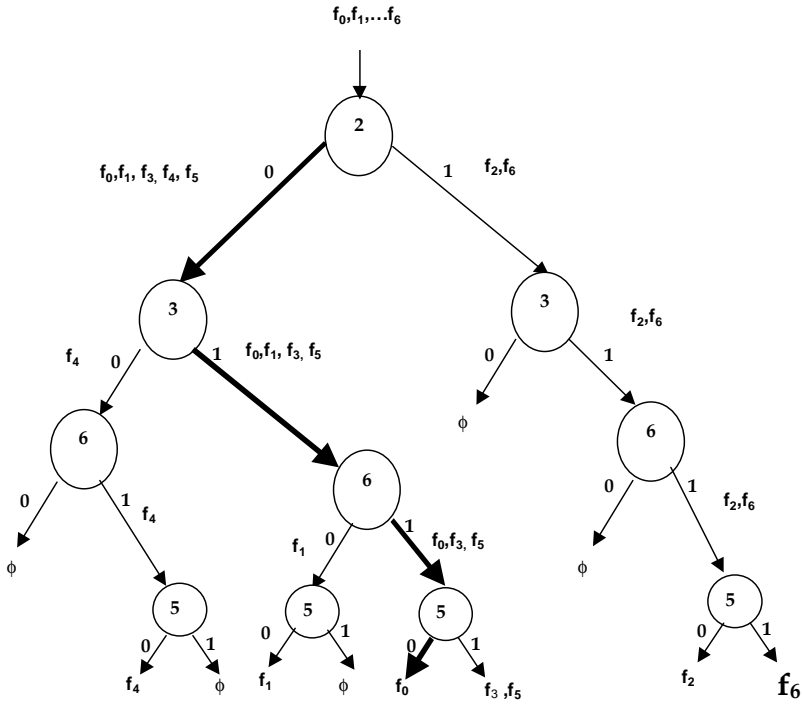

**Figure 1 (Diagnosing Tree for fault Detection)**

## 3.2: Heuristic method

This method consists of two phases. They are

**Phase -1: Construction of the fault dictionary**

**Phase - 2 : Proposed Method for Test minimization**

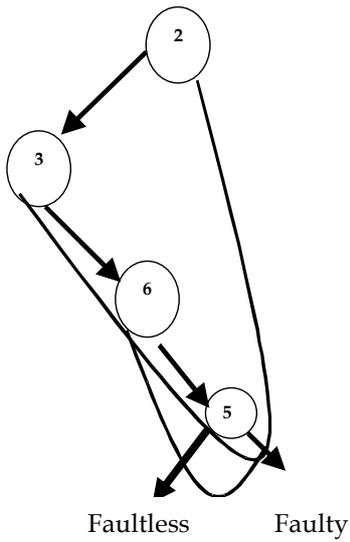
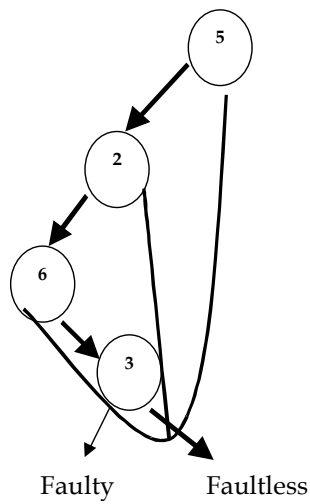

**Figure 2(Simplified tree)**  **Figure 3(another simplified tree)**

**Phase -1: Construction of the fault dictionary**

If $x_1, x_2, \ldots, x_n$ are the input variables to a single output circuit whose fault-free(correct) output is $z = z(x_1, \ldots, x_n)$ and $z^{\alpha_1}, z^{\alpha_2}, \ldots z^{\alpha_i}$ are the erroneous outputs, each corresponding to one of the possible faults $\alpha_1, \alpha_2, \ldots \alpha_i$, a multiple-output table of the combinations may be obtained. This is called a fault dictionary, F in Table I.

Table I Fault Dictionary

| Row Number | $x_1$ | $x_2$ | … | $x_n$ | $z$ | $z^{\alpha}_1$ | $z^{\alpha}_2$ | … | $z^{\alpha}_i$ |
|---|---|---|---|---|---|---|---|---|---|
| 0 | 0 | 0 | … | 0 | 0 | 1 | 0 | … | 0 |
| 1 | 0 | 0 | … | 1 | 1 | 1 | 0 | … | 1 |
| . | . | . | . | . | . | . | . | . | . |
| . | . | . | . | . | . | . | . | . | . |
| . | . | . | . | . | . | . | . | . | . |
| $2^n-1$ | 1 | 1 | … | 1 | 0 | 0 | 0 | … | 1 |

**Phase -2 : Proposed Method for Test minimization**

This method adopts heuristic approach for test minimization. After two test inputs have been applied, the four partial test schedules that should follow may be all different in content and length, and so on for successive test inputs. The objective of this method is to find test schedule with minimum number of levels in the diagnosing tree. The following factors influence the method adopted for finding the minimal test set:

1. The tests chosen must not be confined to any particular given set of tests. They must be chosen from the rows of the fault table.
2. The construction of an adaptive schedule of tests for fault location with a minimal number of levels needs at least $N_L$ tests.
3. At each step the test that will distinguish between the largest number of faults not already distinguished should be chosen.

One way to select the appropriate row (test) at each step of procedure is to try all possible remaining rows(tests). For even a small fault table, however, the number of possible graph labeling that must be tried to determine the minimal number of levels is astronomical. This approach is therefore impractical.

Let $W_{i0}$ and $W_{i1}$ denote the numbers of 0's and 1's in row i, respectively. A simple heuristic method for finding a nearly minimal adaptive-scheduled fault-location experiment is:

Select row i, which maximizes the number of (0,1) pairs between digits in that row, that is, which maximizes the expression

$$R_i = W_{i0} W_{i1}$$

This number is optimized if the row that has the most nearly equal distribution of 0's and 1's, [i.e. the row (or one of the subset of rows) for which $|W_{io}-W_{i1}|$ is minimal] is selected. The use of this criterion appears to work very well for many problems.

## 4. COMPLETE ALGORITHM FOR TEST MINIMIZATION

This algorithm accepts the given expression in sum of product form, produces fault table, detects faults and outputs essential tests after eliminating redundancy, if any. The pseudocode for proposed Algorithm is given below:

Start

    Initialize a binary tree data structure.
**Step 1**: Read Boolean Expersion(Sum of products form)
**Step 2**: Substitute binary values for all the symbols and find *s-a-0* and *s-a-1* faults
**step 3:** Eliminate duplicate faults
**step 4:** Create fault table by substituting all binary combinations *n* symbols in given experssion
**step 5:** Eliminate duplicate rows and columns in fault table.
**Repeat**
    **Step 6:** Find sum of 1's and sum of 0's in each row
    **Step 7:** Find difference between sums of 0's and 1's
    **Step 8:** Find minimum of the difference.
    **Step 9:** Select the row number with minimum difference as essential number.
    **Step 10**: Split the matrix into two sub matrices one containing 0's and another containing 1's based on essential test row.
    **Step 11**: Store the column numbers in each of the sub matrix as left and right child of the binary tree.
    **Step 12:** Eliminate the selected row number from further analysis
**Until** the left and right child of the binary tree contains a single column number.
**Step 13:** Eliminate redundant test numbers.
**Step 14:** Output essential test for the given circuit.

Stop

## 5. RESULTS AND DISCUSSION

This algorithm is implemented in C language and experimented and tested with different simple combinational circuits. It finds essential test numbers, for the given circuits. The results are tabulated in table II. The following inferences are made on observation of the table II. Execution time of the algorithm increases with the increase in number of gates. Achievement of 99% reduction is due to the fact that the large number of tests find the same faults. Hence, they are grouped and one among them is retained and the rest are eliminated. A diagnosing tree created occupies less storage space.

Part of the algorithm (section 4) proposed in this research work uses the binary tree structure to identify test numbers covering more than one fault and eliminates redundant tests to be performed. Hence, they are grouped and one among them is retained and the rest are eliminated. This method suffers by keeping a very large fault detection and location table. This will increase the memory requirement of the data structures used. The increase in memory requirement is directly proportional to the total number of inputs to the gates.

Table II Results

| Sl. No | No. of Inputs n | Total No of Tests $2^n$ (a) | No. of faults | Minimized Tests (b) | Execution Time in Seconds | Minimization in Percentage (a-b)/a*100 |
|---|---|---|---|---|---|---|
| 1 | 4 | 16 | 8 | 7 | 0.5 | 50.0 |
| 2 | 6 | 64 | 11 | 10 | 1 | 82.8 |
| 3 | 8 | 256 | 14 | 13 | 2 | 94.9 |
| 4 | 9 | 512 | 14 | 13 | 3 | 97.4 |
| 5 | 10 | 1024 | 17 | 15 | 6 | 98.5 |
| 6 | 11 | 2048 | 17 | 15 | 13 | 99.3 |
| 7 | 12 | 4096 | 20 | 17 | 28 | 99.6 |
| 8 | 13 | 8192 | 22 | 19 | 30 | 99.7 |
| 9 | 14 | 16384 | 25 | 23 | 34 | 99.8 |
| 10 | 15 | 32768 | 29 | 24 | 35 | 99.9 |

## 6. CONCLUSION

To minimize the test generation problem in simple two stage combinational circuits, a new heuristic algorithm has been developed. This algorithm is implemented and tested with different combinational circuits. From the implementation it was observed that execution time of the algorithm increases with the increase in number of gates. Test minimization varies from 50% to 99% with the lowest one corresponding to a circuit with four gates. In case of complex circuits, number of faults are naturally more. Test minimization percentage reduces in those cases. This Algorithm requires a very large fault table which is to be constructed and provides optimal solution. This procedure is quite simple and easy to apply. The drawback of this method is that it requires a large amount of computer storage space to store the fault table. The next phase of the research work is extended to develop suitable Heuristic search Algorithm like Genetic Algorithm to overcome the difficulties of proposed method. This work may be extended to very large scale integration (VLSI) like benchmark circuits.

**Authors**

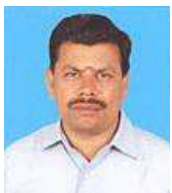
**Prof. Dr K.Kuppusamy** is working as an Associate Professor, Department of Computer Science and Engineering, Alagappa University, Karaikukdi, Tamilnadu, India. He has received his Ph.D in Computer Science and Engineering from Alagappa University, Karaikudi, Tamilnadu in the year 2007. He has published many papers in International Journals the presented in National and International conferences. His areas of research interests include Information/Network Security, Algorithms, Neural Networks, Fault Tolerant Computing, Software Engineering & Testing and Operational Research.

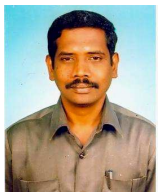
**Prof. T.Meyyappan** currently, Associate Professor, Department of Computer Science and Engineering, Alagappa University, Karaikudi, TamilNadu. He has submitted his Ph.D. thesis in Computer Science in May 2010 and published a number of research papers in National and International journals and conferences. He has developed Software packages for Examination, Admission Processing and official Website of Alagappa University. His research areas include Operational Research, Digital Image Processing, Fault Tolerant computing, Network security and Data Mining.

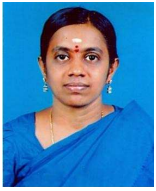 **SM.Thamarai** received her Diploma in Electronics and Communication Engineering in Department of Technical Education, TamilNadu in 1989 and her B.C.A., M.Sc. (First Rank holder and Gold Medalist), M.Phil. (First Rank holder) degrees in Computer Science(1998-2005) from Alagappa University. She has published two research papers in International Journals and presented papers in Six National and one International Conferences. Her current research interests are in Operational Research and Fault Tolerant Computing. She is currently pursuing her Ph.D. in Alagappa University, Karaikudi, TamilNadu.